\documentclass[a4paper,11pt,epsf]{article}

\begin{document} 
\hrule 
\leftline{}
\leftline{Chiba Univ. Preprint
          \hfill   \hbox{\bf CHIBA-EP-131}}
\leftline{\hfill   \hbox{hep-th/0110004}}
\leftline{\hfill   \hbox{September 2001}}
\vskip 5pt
\hrule 
\vskip 0.5cm

\centerline{\Large\bf 
A confining string theory derived from  QCD%
\footnote{Contribution to 10th Tohwa international Symposium on  ``String Theory" held at Tohwa University, Fukuoka, Japan, 3-7 July 2001.} 
}
\vskip 0.5cm

\centerline{{\bf 
Kei-Ichi Kondo
}}  
\vskip 0.5cm
\begin{description}
\item[]{\it \centerline{ 
Department of Physics, Faculty of Science, 
Chiba University,  Chiba 263-8522, Japan}
}
\end{description}

\begin{abstract}
 We report an attempt of deriving a string representation of QCD based on a novel vacuum condensate of mass dimension 2.
\end{abstract}

This work is an attempt to derive a string representation \cite{Polyakov96} of quantum chromodynamics (QCD).  This problem has already been studied by many physicists for the last two decades.  However, the achievement is not yet reached to a satisfactory level.  In my opinion, the main obstacle comes from the fact that the mechanism of mass generation is not made clear in the following sense.  In the classical level, the Yang-Mills theory and QCD with massless quarks have no intrinsic scale, since all the parameters characterizing the Lagrangian are dimensionless, e.g., the coupling constant $g$ and the numbers of colors $N_c$ and flavors $N_f$.  
In the quantum level, however, the dimensional transmutation is expected to occur and the quantum theory has a mass scale $\Lambda$ which is a renormalization group (RG) invariant.  
The non-vanishing constituent mass of the quark has been understood as a result of dynamical mass generation associated with the spontanesous breaking of chiral symmetry, even if the original Lagragian has zero bare mass for the quark.  On the contrary, the origion of the gluon mass is not yet understood in the similar way.  

\par
This work is based on a novel vacuum condensate of mass dimension 2
which has recently been proposed, independently by several authors \cite{Schaden99,KS00,Boucaudetal00,GSZ01}.  
We begins with the BRST formulation of the gauge theory.  In the most general form \cite{Kondo01}
$
 \mathcal{O} := {1 \over \Omega^{(D)}} \int d^D x \ {\rm tr} \left[ 
 {1 \over 2} A_\mu(x) \cdot A_\mu(x) + \alpha i \bar{C}(x) \cdot C(x) 
\right] ,
$
where $\Omega^{(D)}$ is the volume of the $D$-dimensional spacetime.
It has been shown \cite{Kondo01} that the composite operator $\mathcal{O}$ is invariant under the BRST and anti-BRST transformations in the manifestly Lorentz covariant gauge, especially in the most general
 Lorentz gauge \cite{BT82} and the Maximal Abelian (MA) gauge \cite{tHooft81,KondoI,KondoII,KondoIV}.  
Here the trace is taken over the broken generators of the Lie algebra $\mathcal{G}$ of the original group $G$ (broken by the local gauge fixing condition from $G$ to $H$), i.e, $G$ itself for the Lorentz gauge and $G/H$ for the  MA gauge where $H$ is the maximal torus group of $G$.  
Especially, in the limit $\alpha \rightarrow 0$ (which we call the Landau gauge), the composite operator becomes gauge invariant, since the contribution from the ghost and anti-ghost disappears.  
The vacuum condensate includes the ghost condensation proposed in the MA gauge \cite{Schaden99,KS00} and reduces to the gluon condensation proposed recently by several authors \cite{Boucaudetal00,GSZ01}.
In the paper \cite{Kondo01}, we have discussed that such a vacuum condensate may provide the essential ingredients of mass gap and quark confinement in the Yang-Mills theory.  
\par
The first attempt of obtaining the string representation of QCD based on the above proposal  was given for the Yang-Mills theory in the MA gauge, see \cite{KondoI,Kondo00}. (The similar attempt has been reported \cite{Ellwanger98}.)
Now this approach can be extended to any Lorentz covariant gauge including the Lorentz gauge as well as the MA gauge.
In order to obtain the equivalent dual theory of the Yang-Mills theory with an insertion of the Wilson loop operator, we have introduced an antisymmetric (Abelian) tensor field as an auxiliary field of an gauge invariant antisymmetric tensor field $f_{\mu\nu}^\xi$ which appears in the non-Abelian Stokes theorem for the Wilson loop.   
The obtained antisymmetric tensor gauge theory can be converted to a dual Ginzburg-Landau theory, a magnetic monopole current theory and a confining string theory, as suggested \cite{KondoI} and demonstrated in \cite{Kondo00}.  
The details will be given in a subsequent paper in preparation.

\end{document}